# Effect of substrate temperature on the growth of Nb$_3$Sn film on Nb by multilayer sputtering

Md Nizam Sayeed[1,2], Uttar Pudasaini[3], Charles E. Reece[3], Grigory V. Eremeev[4], and Hani E. Elsayed-Ali[1,2*]

[1]*Department of Electrical and Computer Engineering, Old Dominion University, Norfolk, VA 23529, USA*

[2]*Applied Research Center, Old Dominion University, 12050 Jefferson Avenue, Newport News, VA 23606, USA*

[3]*Thomas Jefferson National Accelerator Facility, Newport News, VA 23606, USA*

[4]*Fermi National Accelerator Laboratory, Batavia, IL 60510, USA*

**Abstract**

We report on the fabrication of niobium tin (Nb$_3$Sn) films by multilayer sequential sputtering on niobium at substrate temperatures ranging from room temperature to 250 °C. The multilayers were then annealed inside a separate vacuum furnace at 950 °C for 3 h. The material properties of the films were characterized by X-ray diffraction, scanning electron microscopy, energy-dispersive X-ray spectroscopy, atomic force microscopy, and transmission electron microscopy. The superconducting properties of the films were studied by four-point probe resistivity measurements from room temperature to below the superconducting critical temperature $T_c$. The highest film $T_c$ was 17.76 K, obtained when the multilayers were deposited at room temperature. A superconducting Nb$_3$Sn thin film with a smooth surface morphology and $T_c$ of 17.58 K was obtained on the film deposited at a substrate temperature of 250 °C.

## 1. Introduction

Nb$_3$Sn is an intermetallic, type II superconductor of A15 crystal structure that, in the bulk form, has a superconducting critical temperature $T_c$ of ~18 K, a superheating field $H_{sh}$ of ~400 mT, and an upper critical magnetic field $\mu_0 H_{c2}$ of 30 T [1,2]. The high $T_c$ and high $\mu_0 H_{c2}$ of Nb$_3$Sn enable its use for superconducting wires to replace the Nb-Ti alloy ($T_c$ = 9.8 K, $\mu_0 H_{c2}$ = 14.5 T) [3]. Nb$_3$Sn is also considered a potential candidate for superconducting radiofrequency (SRF) accelerator cavities. Current radiofrequency accelerating fields of bulk niobium cavities are close to the superheating field $H_{sh}$ of 200 mT, which limits the accelerating gradient of the cavities to ~50 MV/m [4,5]. Niobium SRF cavities are chemically polished to remove defects that could prevent the cavity from sustaining magnetic fields close to $H_{sh}$ of Nb. Efforts are ongoing to reduce imperfections on the Nb$_3$Sn surfaces to raise their superheating critical field $H_{sh}$ closer to that of Nb [6,7]. Since the quality factor $Q_0$ of a cavity is inversely proportional to its surface resistance, Nb$_3$Sn which has a lower surface resistance than Nb and higher $T_c$ and $H_{sh}$ is considered a potential alternative to Nb for SRF cavities. The higher $T_c$ enables the SRF cavity to be operated at 4.2 K instead of the 2 K often needed for Nb cavities, which can significantly reduce the accelerator operating cost. However, bulk Nb$_3$Sn cannot be used to fabricate a cavity due to its brittle nature. Instead, Nb$_3$Sn-coated Nb cavities are considered. Nb$_3$Sn films were grown inside Nb cavities by Sn vapor diffusion [8]. Nb$_3$Sn vapor diffusion-coated 1.3 GHz Nb single-cell cavities have achieved $Q_0 \geq 2 \times 10^{10}$ at 4 K with an accelerating gradient up to 15 MV/m, as reported by Jefferson Lab [9], and a maximum accelerating gradient of 22.5 MV/m at 4 K, as reported by Fermilab [6]. At Jefferson Lab, a five-cell cavity achieved a maximum accelerating gradient of 15 MV/m at 4 K [10] and a low field $Q_0$ of $3 \times 10^{10}$ [11]. Nb$_3$Sn-coated high-frequency cavities tested at Cornell also achieved a high $Q_0$ at 4 K (~$8 \times 10^9$ for 2.6 GHz cavities, and ~$2 \times 10^9$ for 3.9 GHz cavities) [12].

In addition to the vapor diffusion method, Nb$_3$Sn films were deposited on different substrates by electrochemical synthesis [3,13], dipping Nb into liquid Sn [14], evaporation [15,16], sputtering Nb on hot bronze [17], and magnetron sputtering [18-31]. Magnetron sputtering is a promising alternative to vapor diffusion for coating Nb$_3$Sn inside Nb or Cu cavities due to its ability to produce homogenous films on substrates of different shapes [23]. Nb$_3$Sn is fabricated by magnetron sputtering from a single stoichiometric Nb$_3$Sn target [18-25] or using separate Nb and Sn sputter targets [26-31]. Two different processes can be applied to fabricate Nb$_3$Sn films using separate Nb and Sn targets: multilayer sputtering of Nb and Sn followed by annealing, and co-sputtering of Nb and Sn simultaneously. Both methods have been used to fabricate high-quality superconducting Nb$_3$Sn films with a $T_c$ ranging from 15 to 18 K, depending on the substrate material and deposition parameters.

We studied the fabrication of Nb$_3$Sn films on Nb substrates by multilayer sequential sputtering. The goal of the present study is to identify conditions that yield superconducting Nb$_3$Sn film which can be applied for coating the inside surface of Nb SRF cavities. Co-sputtering of Nb and Sn requires inserting both targets inside the SRF cavity and controlling the Nb and Sn ratios deposited over curved surfaces, which complicates the process for a small cavity. In multilayer sputtering, the cylindrical targets can be inserted in the SRF cavity separately allowing for each layer deposition to be controlled independently. The Nb-Sn multilayers were annealed at 950 °C for 3 h to form the A15 Nb$_3$Sn phase. The effects of the thickness of the multilayers, thickness of the Nb buffer layer deposited on the Nb substrate before depositing the multilayers, and the annealing temperature were studied previously [25-31]. Although some of the Nb$_3$Sn films reached a $T_c$ of 17.93 K [29], the surface of the film had a large distribution of voids, which can affect RF superconducting performance. Here, we examined the properties of the Nb$_3$Sn film with varied deposition substrate temperatures. Nb-Sn multilayers were deposited at different substrate temperatures starting from room temperature (RT) up to 250 °C. The multilayers were then annealed to fabricate the Nb$_3$Sn film. The surface morphology was observed to vary significantly with the increase in the substrate temperature. Nb$_3$Sn with a $T_c$ of 17.58 K and a smooth surface with low void density was deposited on Nb.

2. Methods

Multilayers of Nb and Sn were deposited on Nb and sapphire substrates using an AJA ATC Orion-5 commercial magnetron sputter coater. The Nb and Sn films were sputtered from 99.95% pure Nb and 99.999% pure Sn targets with 2″ diameters. The substrate holder is located at the top of the deposition chamber and the distance between the target and the substrate holder is ~10 cm. The substrate holder was rotated at 30 rpm during the coating to achieve homogenous film thickness. The substrate temperature was varied using a resistive heater placed at the back of the substrate holder. The Nb substrates were prepared from Nb sheets (*RRR* ~250) used for SRF cavity fabrication. The substrates were prepared by applying the exact processing steps used for cavity fabrication, namely, cutting 1 cm × 1 cm Nb substrates by wire electro-discharge, buffered chemical polishing (BCP 1:1:1 to remove 100 μm), baking at 800 °C for 2 h in a vacuum furnace with background pressure at or below 10$^{-3}$ Pa, and a second BCP to remove more 25 μm from the surface. The substrates were 430-μm thick double-side polished sapphire with C-M orientation from University Wafers Inc. The films deposited on the sapphire substrates were used to study the superconducting properties. Before deposition, the substrates were cleaned with ethanol and isopropanol.

Before deposition, the sputter chamber was evacuated to ~$10^{-5}$ Pa and the deposition was carried out at an Ar background pressure of $4 \times 10^{-1}$ Pa. The flow rate of the Ar gas (20 SCCM) was controlled by a mass flow controller. Initially, a 200 nm thick Nb buffer layer was deposited on the substrate. Then, multilayers of Nb and Sn were deposited on top of the buffer layer. The thickness of the Nb and the Sn layers were 20 and 10 nm, respectively. This sequence of Nb and Sn layer deposition was repeated 34 times to deposit a multilayer film about 1 µm thick. The final deposited layer was 20 nm Nb, which acts as a barrier to reduce Sn evaporation during annealing. The top Nb layer reacted with diffusing Sn during post-deposition annealing to form $Nb_3Sn$. The deposition rate for both the Nb and Sn was 1 Å/s. The deposition of the multilayers was conducted at substrate temperatures of RT, 100, 150, and 250 °C. All multilayers were post-annealed inside a separate vacuum furnace at 950 °C for 3 h. This annealing temperature and time were chosen based on our previous experiments using different annealing temperatures and annealing times [28]. Before annealing, the furnace was evacuated to ~$10^{-3}$ Pa. The temperature was increased at a ramp rate of 12 °C/min. After annealing, the heater was turned off while maintaining the vacuum till the sample temperature reached below 50 °C.

The film structures were studied by X-ray diffraction (XRD) of the films for 2θ between 10 and 90° using a Rigaku Miniflex II X-ray diffractometer by generating CuKα radiation ($\lambda$ =1.54056 Å) from a copper tube operating at 30 kV and 15 mA. The surface morphologies of the films were observed by a Hitachi S-4700 field emission scanning electron microscopy (SEM) and a Jeol-JSM 6060 LV SEM. The elemental compositions of the films were performed by Noran 6 energy dispersive X-ray spectroscopy (EDS) detector using 15 keV accelerating voltage. EDS was performed on the 1.2 µm$^2$ area at a working distance of 10 mm. The high-resolution transmission electron microscopy (TEM) images were obtained by a ThermoFisher Titan 80-300 probe aberration-corrected scanning TEM using an operating voltage of 200 kV. The cross-sectional films for TEM were prepared by focused ion beam sectioning using Thermofisher Quanta 3D FEG. The surface texture and surface roughness of the films were obtained from a Digital Instrument atomic force microscope (AFM) operated in the tapping mode. The superconducting properties of the films deposited on sapphire substrates were characterized by the resistance versus temperature data obtained from the four-point probe measurement. The samples were loaded into an isothermal liquid-helium-cooled cryostat with a temperature resolution better than 50 mK [32].

## 3. Results and Discussion

*3.1. Structural and Morphological Properties*

Figures 1(a) and 1(b) show the XRD peaks of as-deposited and annealed films, respectively. The as-deposited films have diffraction peaks from Nb for all deposition conditions. All the annealed films have polycrystalline $Nb_3Sn$ with diffraction peaks identified as the 110, 200, 210, 211, 222, 320, 321, 400, 420, 421, and 332 orders. The XRD peak intensity is compared with the $Nb_3Sn$ PDF card (PDF Card No.: 00-017-0909 Quality: I) and the relative peak intensity is about the same as the standard data for the film deposited at room temperature. The most intense peak is found for the 210 diffraction order. As the substrate temperature increased, the relative intensity of 200 and 320 diffraction orders increased gradually. At the substrate temperature of 250 °C, the intensity of 320 diffraction surpassed the intensity of 321 diffraction order. The increased intensity of the $Nb_3Sn$ diffraction orders with increased substrate temperature confirms the improved film crystallinity at the higher deposition temperatures. The crystallite size was calculated from the width of the 200, 210, and 211 diffraction orders using the Scherrer equation [33]:

$$D = \frac{0.97\lambda}{B\cos\theta} \quad (1)$$

where $\lambda$ is the X-ray wavelength, $\theta$ is the Bragg angle, and $B$ is the full width at half-maximum (FWHM) of the diffraction peak. The crystallite sizes, calculated from Eq. (1), for films deposited at different substrate temperatures, are shown in Figure 2. For all $Nb_3Sn$ diffraction orders, the crystallite size increased with increasing substrate temperature. The crystallite size calculated from the 200 diffraction order changed from ~39 to ~80 nm when the substrate temperature was raised from RT to 250 °C.

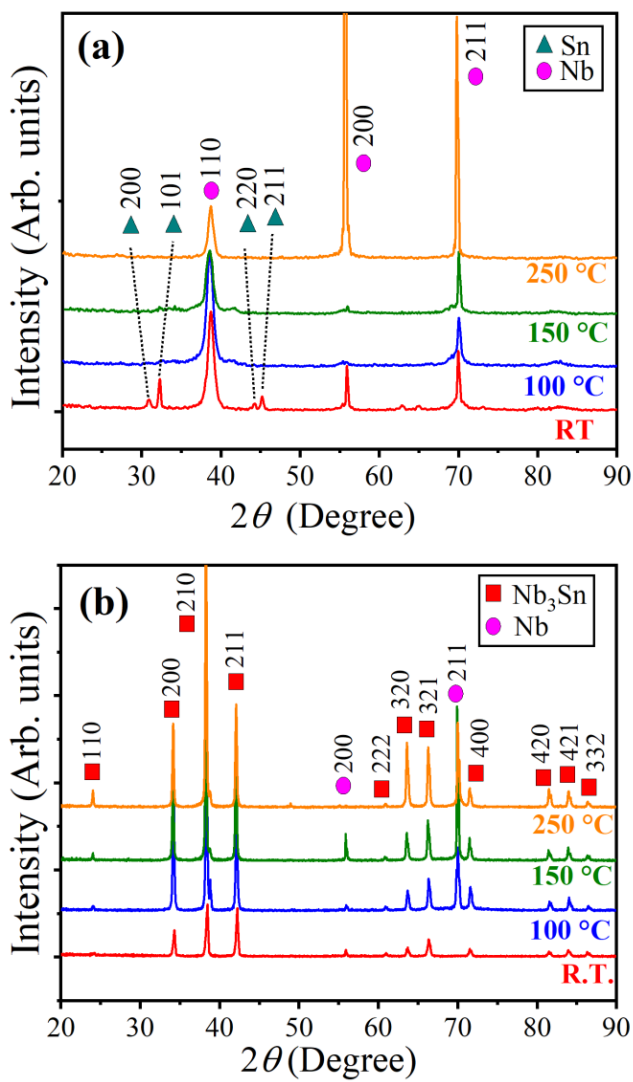

Figure 1. X-ray diffraction peaks of (a) as-deposited, and (b) annealed films sputtered at a substrate temperature of RT, 100, 150, and 250 °C.

The surface morphology was highly dependent on the substrate temperature during the deposition of the multilayers. Figure 3 shows the SEM surface images of as-deposited films grown at different substrate temperatures. The film coated at RT has a fine grain structure. For substrate temperature of 100 °C, some grains coalesced and created grain clusters. When the substrate temperature was further increased to 150 °C, the clustered regions separated from each other, leaving flat surface areas between the clusters. These flat areas increased, and the cluster size decreased when the substrate temperature was further increased to 250 °C. The clusters are formed by Sn grains coalescing at high temperatures. EDS mapping of the films in Figure 4 confirmed the abundance of Sn near the clusters.

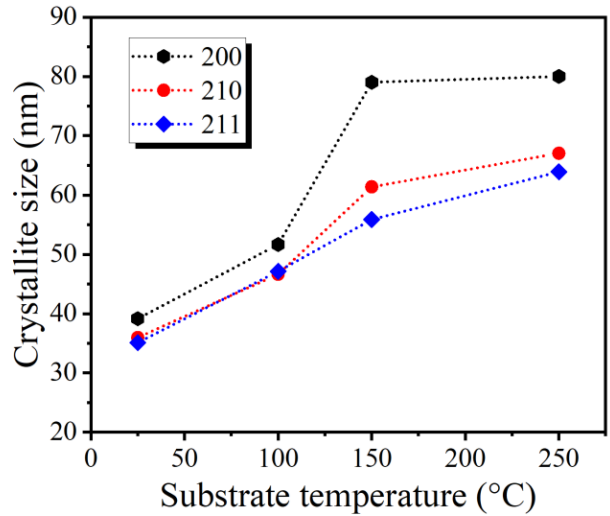

Figure 2. The Crystallite size of the Nb$_3$Sn was calculated from the (200), (210), and (211) diffraction orders for different substrate temperatures.

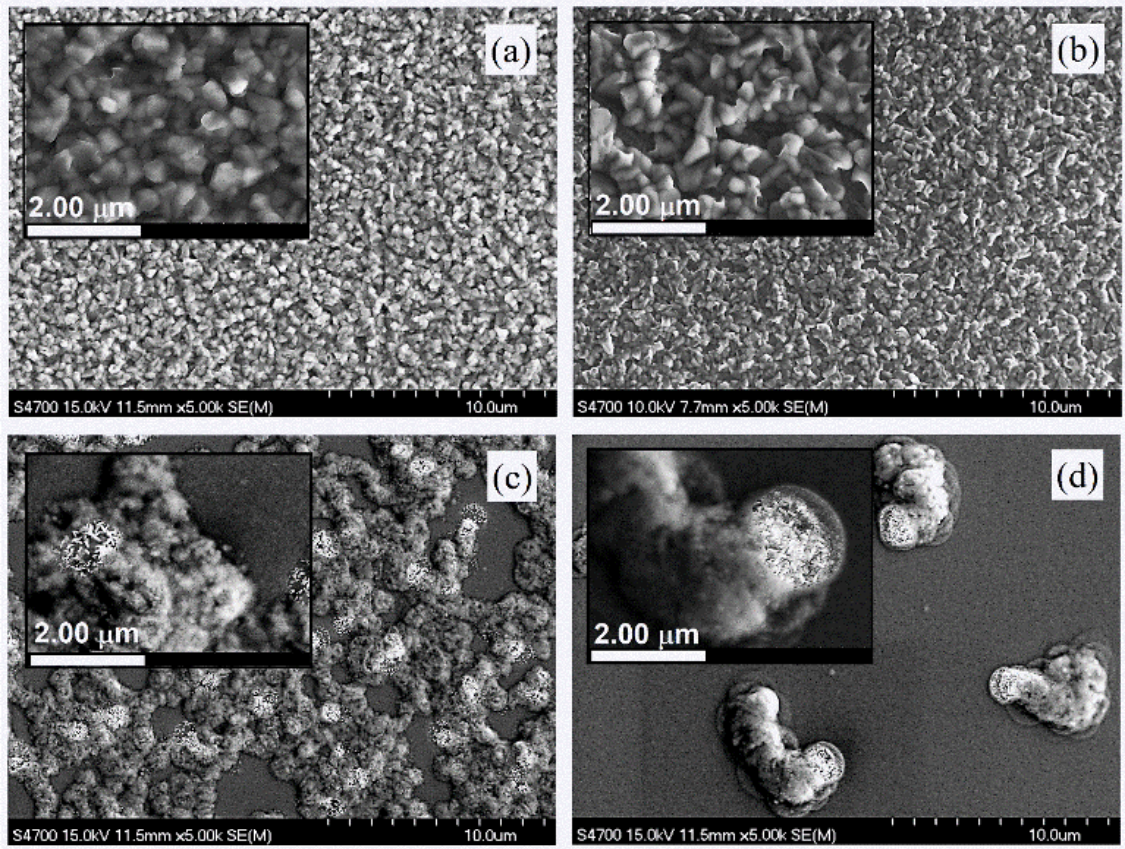

Figure 3. SEM images of the as-deposited Nb-Sn films on Nb substrates for different temperatures: (a) RT, (b) 100 °C, (c) 150 °C, and (d) 250 °C. The total thickness of the films is ~1 µm.

Figure 4(a) shows an SEM image of Nb-Sn multilayered films deposited at 250 °C. The arrow indicates the location of the EDS line-scan for the elemental mapping shown in Figure 4(b). The EDS line-scan of the surface confirms the increased Sn-concentration across the clusters. The enhanced intensity near the clusters in Figure 4(c) shows the abundance of Sn on the clusters. Also, the EDS mapping of Nb shown in Figure 4(d) confirms Nb deficiency in the same region.

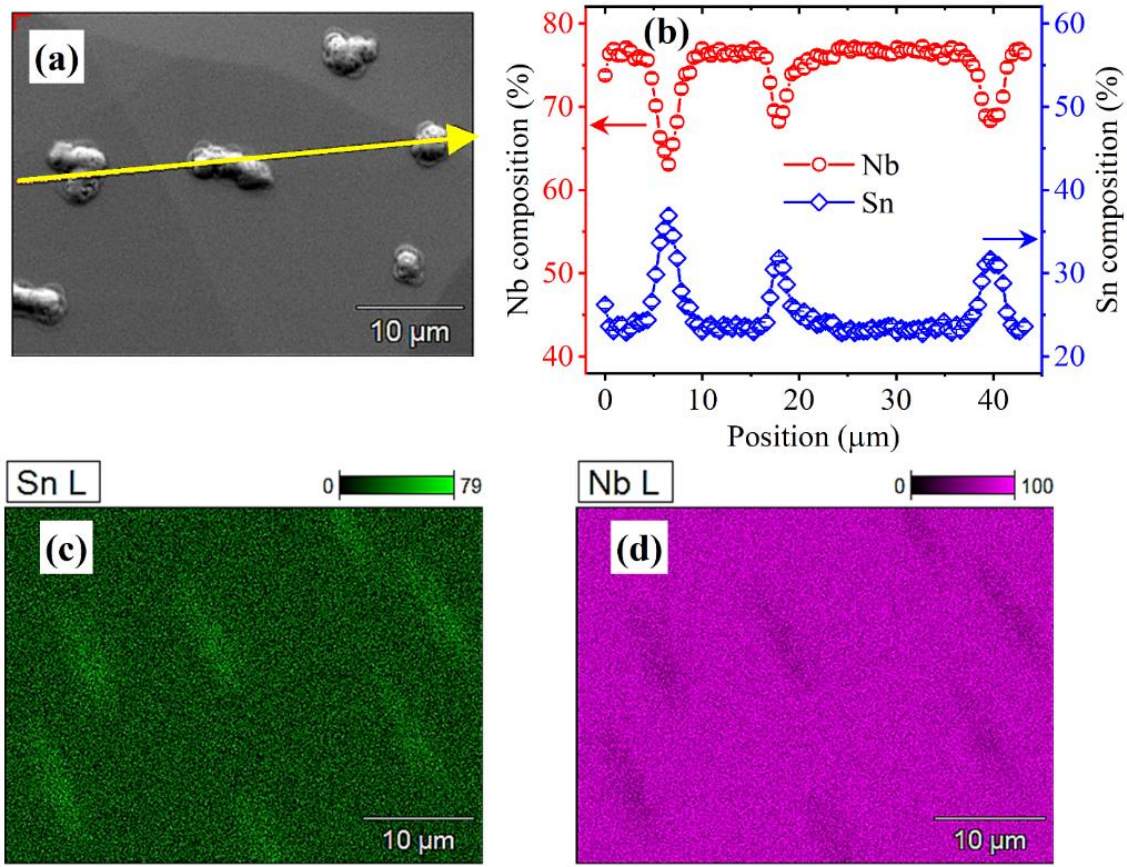

Figure 4. (a) SEM image of Nb-Sn multilayered films deposited at 250 °C. The arrow shows the location of the EDS line scan. (b) Corresponding elemental composition from the EDS line-scan. (c) Elemental mapping of Sn. (d) Elemental mapping of Nb.

The SEM images of the annealed films are shown in Figure 5. The insets show the magnified images of the surface. The film coated at RT, shown in Figure 5(a), had islands with an average size of ~200 nm and shows some voids distributed throughout the surface. The films coated at 100

°C, shown in Figure 5(b), had fewer voids than in the film deposited at RT. The average size of the islands forming the clusters was smaller than the islands in the flat areas between the clusters. Similar morphology was also observed for the annealed films coated at 150 and 250 °C, shown in Figures 5(c) and 5(d), respectively. Figures 5(e) and 5(f) show the SEM images at locations 1 (where the voids are found) and 2 (void-free location), respectively, obtained from Figure 5(d).

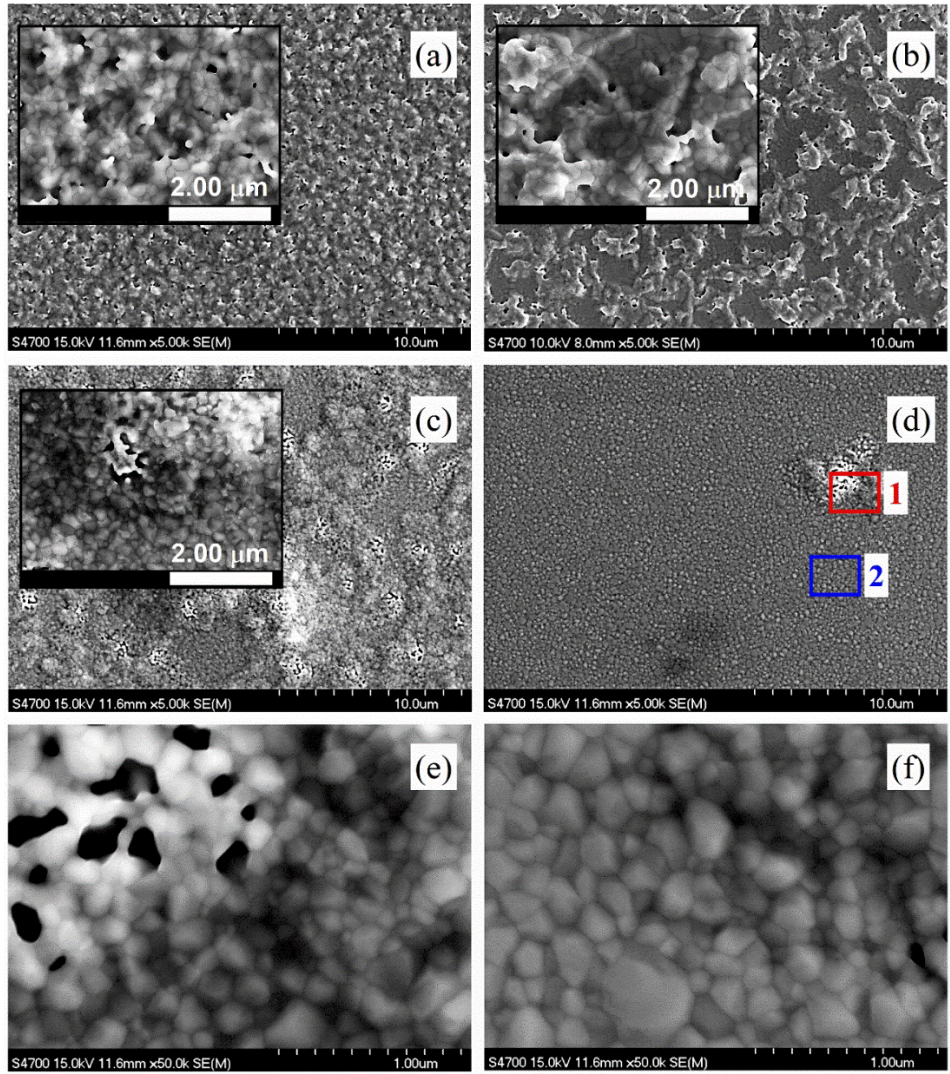

Figure 5. SEM images of the annealed $Nb_3Sn$ films on Nb substrates deposited at different temperatures: (a) RT, (b) 100 °C, (c) 150 °C, and (d) 250 °C. (e) Magnified image of "area 1 of (d)", and (f) Magnified image of "area 2 of (d)".

The Sn composition of the films was measured by EDS over an area of 1.2 mm$^2$. The results are summarized in Table 1. The atomic compositions were derived by averaging the Sn compositions at five different locations on the films. The as-deposited film had ~25% Sn, which was reduced to ~21% when annealed at 950 °C for 3 h.

**Table 1.** Sn composition of the as-deposited and annealed films coated at different temperatures obtained by EDS.

| Deposition temperature (°C) | Sn % As-deposited | Sn % Annealed |
| --- | --- | --- |
| RT | 25.4 ± 0.7 | 21.1 ± 0.3 |
| 100 | 24.6 ± 0.8 | 20.7 ± 0.3 |
| 150 | 25.1 ± 0.4 | 21.2 ± 0.4 |
| 250 | 24.7 ± 0.8 | 21.2 ± 0.4 |

The AFM images recorded for a scan area of 20 µm × 20 µm are shown in Figure 6(a-e). Figure 6(a) shows the AFM image of the Nb substrate before coating and Figure 6(b-e) shows the AFM images of the films deposited at RT, 100, 150, and 250 °C and annealed at 950 °C for 3 h. The root-mean-square (RMS) surface roughness of the films is shown in Table 2. The Nb substrate had RMS roughness of ~16 nm. The roughness of the films increased from ~34 to 51 nm when the substrate temperature was increased from RT to 100 °C. Figure 6(f) shows the line-scan of the films along the arrows indicated in Figure 6(b-e). The maximum peak-to-valley height of the cluster in Figure 6(c) was ~280 nm. At the substrate temperature of 150 °C, the surface had clusters of similar texture separated by small flat areas, as observed in the SEM image in Figure 5(c). The RMS roughness was ~32 nm. Further increase in the substrate temperature increased the area of the separated smooth flat regions, which was accompanied by a small increase in the surface roughness. The clusters had a peak-to-valley height of 50 – 120 nm. Since the surface morphology shown in Figure 6 is measured over an area of 20 µm × 20 µm, the measure of roughness should be considered because of the limited surface area probed.

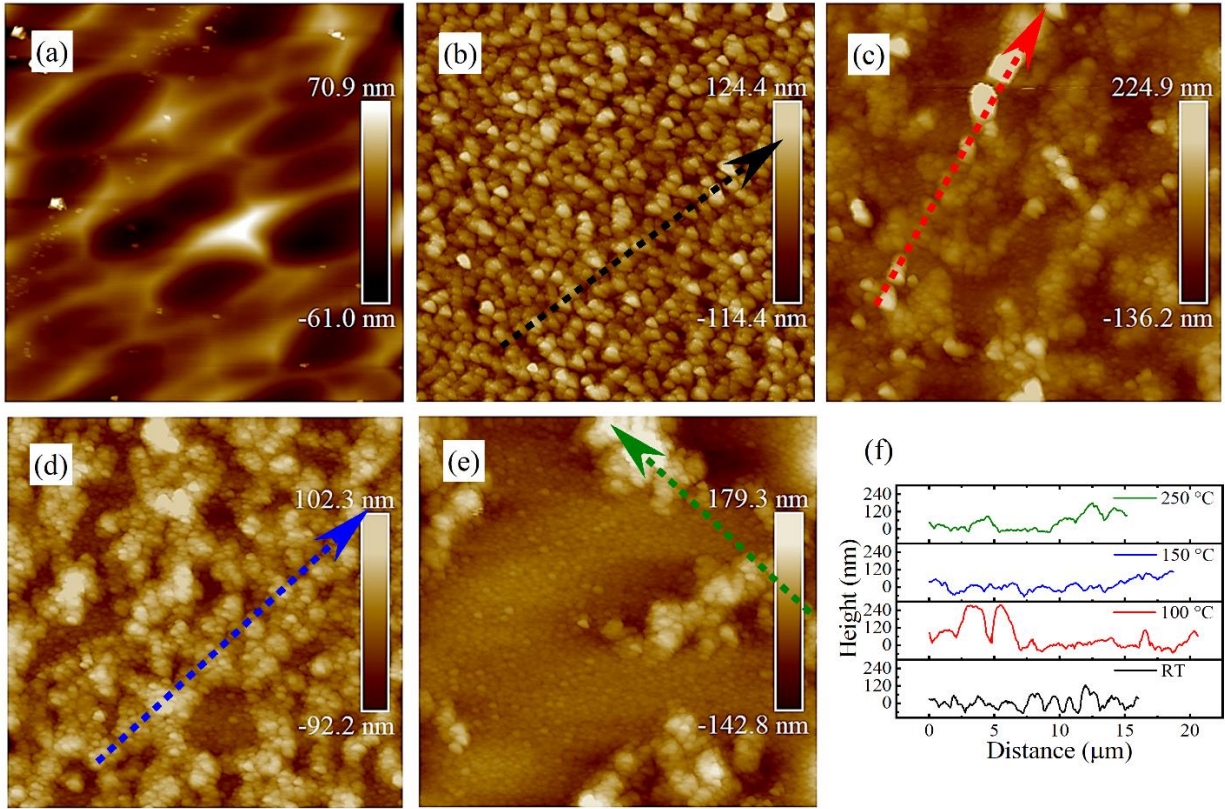

Figure 6. (a) AFM image of Nb substrate, (b-e) AFM images of Nb$_3$Sn films fabricated at a substrate temperature of (b) RT, (c) 100 °C, (d) 150 °C, (e) 250 °C, (f) Line-scans of the films obtained from the AFM images. The thickness of the Nb$_3$Sn films is ~1 μm.

**Table 2.** Root-mean-square (RMS) roughness of the films deposited at different substrate temperatures after annealing at 950 °C for 3 h.

| Deposition temperature (°C) | RMS roughness (nm) |
|---|---|
| RT | 33.6 ± 0.5 |
| 100 | 50.6 ± 4.8 |
| 150 | 31.7 ± 3.6 |
| 250 | 37.9 ± 0.7 |

To better understand the development of the voids in the Nb$_3$Sn film, the cross-sections of the as-deposited and annealed films deposited at RT and 250 °C were examined by TEM, as shown in Figures 7(a) and 7(b), respectively. For the films deposited at RT, the as-deposited film showed the columnar growth of the 200 nm thick Nb buffer layer, multilayers of Nb and Sn, and voids throughout the cross-section. The annealed films also had multiple voids throughout the cross-section. The film deposited at 250 °C had the columnar Nb buffer layer and stacks of multilayers. In contrast to the film deposited at RT, the film deposited at 250 °C did not show any voids between the multilayer stacks, as observed in Figure 7(c). However, large grains with a height of ~500 nm were observed between the multilayer stacks. Annealing the Nb-Sn multilayers that were deposited at 250 °C resulted in an Nb$_3$Sn film without voids between the film grains. For the films deposited at RT and 250 °C, some voids were found at the interface of the Nb substrate and the Nb buffer layer. These voids could have formed due to the Kirkendall effect, in which voids are formed near the intermetallic interface where the volume of one phase increases and the volume for the other phase decreases due to the difference in the diffusion rates of the two metals [34-36].

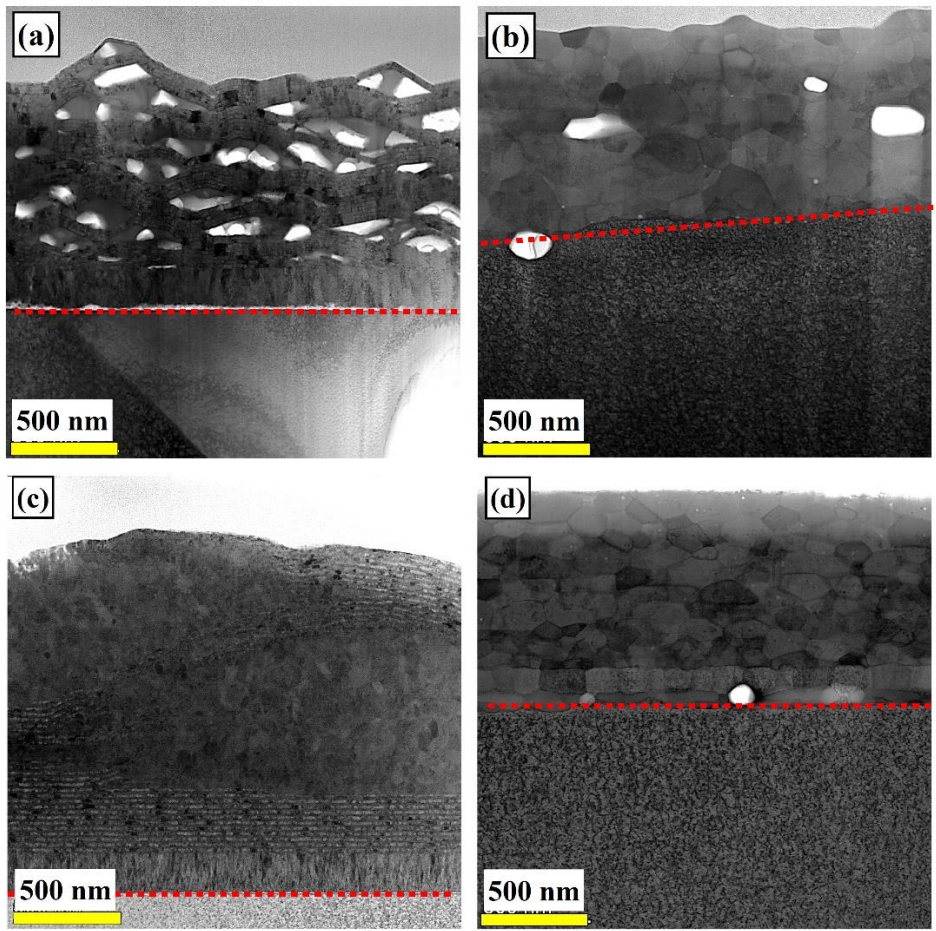

Figure 7. TEM images of the cross-sections of the films: (a) As-deposited film sputtered at RT. (b) The annealed film was deposited at RT. (c) As-deposited film sputtered at a substrate temperature of 250 °C. (d) Annealed film sputtered at a substrate temperature of 250 °C. The dotted red lines indicate the film-substrate interface.

The EDS elemental mappings of the cross-sections of the films are shown in Figure 8. The as-deposited film with the substrate at RT had Sn rich area near the voids. After annealing, $Nb_3Sn$ forms by diffusion of Sn, and the voids remained throughout the cross-section of the film as observed in Figure 8. For the as-deposited film with the substrate at 250 °C, the EDS mapping showed Sn abundance at the large grains, but no voids were observed. Also, no voids were observed in the $Nb_3Sn$ film after annealing. Therefore, diffusion of Sn plays a major role in the formation of voids in $Nb_3Sn$ growth by multilayer sputtering.

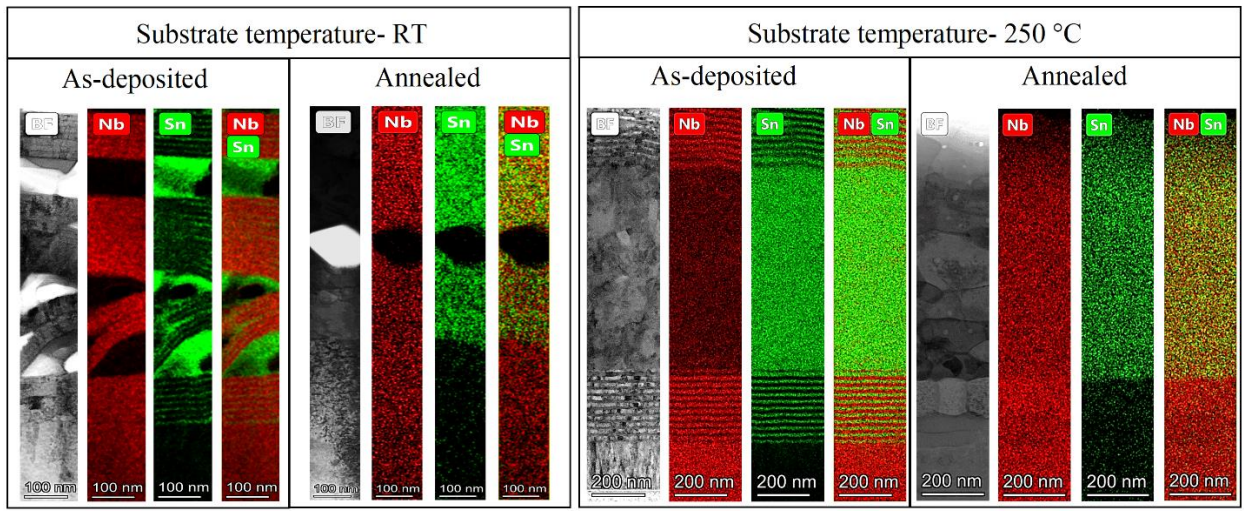

Figure 8. TEM EDS mappings of the cross-sections of the as-deposited and annealed films were deposited at room temperature and at 250 °C.

To better understand the development of Sn-rich clusters on the surface during the growth of the multilayers, we deposited different numbers of Nb-Sn multilayers at a substrate temperature of 250 °C. Figure 9 shows the SEM images of the surface of the films consisting of 1, 3, 10, and 34 layers (each layer is 20-nm Nb, 10-nm Sn). Figure 9(a) shows several bright dotted circles on the surface of the Nb-Sn film of only 1 multilayer. The island growth starts from the first Sn layer as. The inset of Figure 9(a) shows bright elongated grains similar to the bright grains observed in Figure 3(c-d). Increasing the number of multilayers enhances the density and size of the clusters. It appears that Sn clusters are formed at the early stages of the deposition due to the formation of liquid Sn droplets (the melting point of Sn is 231 °C). With increased multilayer numbers, the size of the spherical-shaped grains increases. The spherical grains are probably formed by the condensation of liquid Sn droplets.

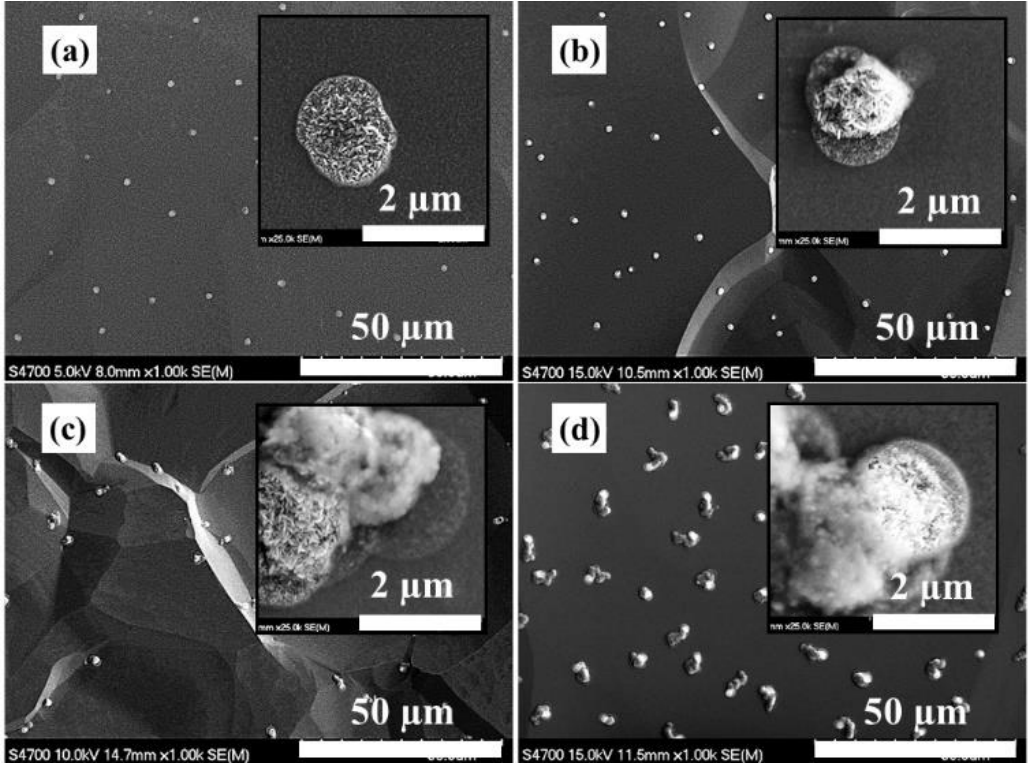

Figure 9. SEM images of the Nb-Sn multilayers deposited at 250 °C with varied multilayer numbers: (a) 1 (b) 3, (c) 10, and (d) 34. The inset shows the magnified images of the surface.

*3.2. Superconducting Properties*

The temperature-dependent resistance of the annealed films measured by four-point probe over the range of 4.5 to 300 K is shown in Figure 10. The resistance decreased as the temperature is lowered, with a sharp resistance decrease observed below 18 K. For all films, $T_c$ was calculated from the average of $T^{90}$ and $T^{10}$ (where $T^{90}$ and $T^{10}$ represent the temperature when the resistance reaches 90% and 10% of the resistance during the transition, respectively). The transition width $\Delta T_c$ was estimated from the temperature difference between the two points $T^{90}$ and $T^{10}$. The residual resistance ratio *RRR* of the film was estimated from the ratio of the resistance at 300 K to the resistance at 20 K. The inset of Figure 10 shows the normalized $R/R_{20K}$ plot near the superconducting transition. The resultant superconducting $T_c$, $\Delta T_c$, and $RRR$ are summarized in Table 3. For all films, $T_c$ varied in a narrow range of 17.58 – 17.76 K, which correspond to the $T_c$ of Nb$_3$Sn. The transition width of the film was observed to increase with increasing the deposition substrate temperature. The film coated at 250 °C had lower $T_c$ and $RRR$ and higher $\Delta T_c$, though the surface of the film had fewer voids. Broader $\Delta T_c$ in Nb$_3$Sn thin films is often found because of non-uniform Sn composition or disorder in the crystalline lattice [23]. Kampwirth et al., observed a broad $\Delta T_c$ of 5.8 K with three different transitions for Nb$_3$Sn films sputtered from a stoichiometric target on a sapphire substrate at a sputtering pressure of 5 mTorr and a substrate temperature of 800 °C [19]. Ilyina et al. reported an average $\Delta T_c$ of 1.7 ± 1.4 K for the Nb$_3$Sn films sputtered from a stoichiometric Nb$_3$Sn target [23]. The $\Delta T_c$ of the multilayered films deposited at RT obtained in this study is below 0.1 K, which is consistent with our previously reported $\Delta T_c$ of multilayered films deposited at RT [28, 29]. The annealed film deposited at 250 °C showed two transitions at 17.4 and 17.8 K, which is due to the Sn composition variation throughout the film.

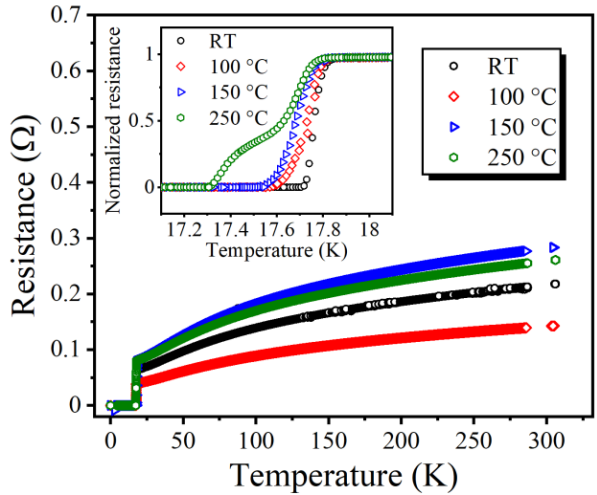

Figure 10. Temperature dependence of the resistances of 1 µm thick Nb$_3$Sn films deposited at different substrate temperatures. The inset of the image shows the plot of normalized resistance ($R/R_{20K}$) with a temperature close to $T_c$.

Table 3. Summary of the superconducting properties of the films.

| Deposition temperature (°C) | $T_c$ (K) | $\Delta T_c$ (K) | RRR |
|---|---|---|---|
| RT | 17.76 | 0.06 | 3.42 |
| 100 | 17.72 | 0.13 | 3.53 |
| 150 | 17.67 | 0.15 | 3.49 |
| 250 | 17.58 | 0.37 | 3.18 |

## 4. Conclusion

Nb$_3$Sn films were fabricated on Nb by sputtering multilayers of 20-nm Nb and 10-nm Sn at RT, 100, 150, and 250 °C followed by annealed at 950 °C for 3 h. Polycrystalline Nb$_3$Sn films were formed with a different surface morphology depending on the substrate temperature. The density of the voids in the films was reduced for the higher substrate temperatures studied, 250 °C. At that temperature, the Nb$_3$Sn film had a $T_c$ = 17.58 K and a surface morphology showing the least voids. However, the non-uniformity in the Sn composition on the films deposited at higher temperatures degraded the superconducting properties. TEM cross-sectional analysis of the films revealed that the voids originate from clusters of Sn embedded in the multilayers. Nb$_3$Sn thin film fabrication by multilayer sputtering can potentially be used for coating Nb SRF cavities with Nb$_3$Sn as multilayer deposition provides good control on the film stoichiometry.

**Conflict of Interest Statement:** The authors have no interest to declare.

**Data availability statement:** All data that support the findings of this study are included in the article.

**Ethics statement:** The work did not involve any human or animal subjects.

**Acknowledgment**

This manuscript is based upon work supported by the U.S. Department of Energy, Office of Science, Office of Nuclear Physics under contract DE-AC05-06OR23177 with Jefferson Science

Associates, including supplemental funding via the DOE Early Career Award to G. Eremeev. The manuscript has been authored by Fermi Research Alliance, LLC under Contract No. DE-AC02-07CH11359 with the U.S. Department of Energy, Office of Science, Office of High Energy Physics. Partial support was provided by the U.S. Department of Energy Contact No. DE-SC0022284. Some of the characterizations were performed at the Core Labs at The College of William and Mary. The authors acknowledge Joshua Spradlin of Jefferson Lab for his help with $T_c$ measurement.